\documentclass[aps,prb,twocolumn,showpacs,preprintnumbers,amsmath,amssymb]{revtex4}
\usepackage{graphicx}
\usepackage{txfonts}

\begin{document}


\title{Anisotropic scattering in angular-dependent magnetoresistance oscillations of quasi-2D and quasi-1D metals: beyond the relaxation-time approximation}
\author{M. F. Smith and Ross McKenzie}
 \email{mfsmith@physics.uq.edu.au}
\affiliation{%
Dept. of Physics, University of Queensland,4072 Brisbane, Queensland, Australia}
\date{\today}

\begin{abstract}
The electrical resistivity for a current moving perpendicular to layers (chains) in quasi-2D (quasi-1D) metals under an applied magnetic field of varying orientation is studied using Boltzmann transport theory.  We consider the simplest non-trivial quasi-2D and quasi-1D Fermi surfaces but allow for an arbitrary elastic collision integral (i.e., a scattering probability with arbitrary dependence on momentum-transfer) and obtain an expression for the resistivity which generalizes that previously found using a single relaxation-time approximation.  The dependence of the resistivity on the angle between the magnetic field and current changes depending on the momentum-dependence of the scattering probability.  So, whereas zero-field intra-layer transport is sensitive only to the momentum-averaged scattering probability (the transport relaxation rate) the resistivity perpendicular to layers measured in a tilted magnetic field provides detailed information about the momentum-dependence of interlayer scattering.  These results help clarify the meaning of the relaxation rate determined from fits of angular-dependent magnetoresistance oscillations (AMRO) experimental data to theoretical expressions.  Furthermore, we suggest how AMRO might be used to probe the dominant scattering mechanism.
\end{abstract}

\pacs{}

\maketitle
\section{Introduction}

Measurements of angle-dependent magnetoresistance oscillations (AMRO) have emerged as a powerful probe of the electronic properties of low-dimensional metals.  When the electric current is perpendicular to the metallic layers (chains) in quasi two(one)-dimensional metals, the measured conductivity
is found to oscillate as a function of the angle $\theta_B$ between the current and applied magnetic field\cite{pipp89, wosn96, lebe86, lebe89, cauf95, blun96, mose99, sing00, kart04, lebe04}.  This dependence, and the weaker but observable variation with the
field-angle $\phi_B$ in the plane perpendicular to the current, depends on the
detailed shape of the Fermi surface and other physical parameters.  Fits to data have been used to extract experimental Fermi surfaces with a resolution comparable to the best available
techniques.\cite{anal07,kawa06, enom06, bali05, kono04, godd04, huss03, kawa03, choi03, sing02, berg00, beck98} Recently, an AMRO study of overdoped cuprates revealed an anisotropic scattering rate with a magnitude that increases monotonically with temperature and scales with the superconducting
transition temperature\cite{abde07,kenn07,abde06}.  Microwave conductivity measurements under a strong field of variable direction provide additional insight, and are being used in Fermiology to augment the zero-frequency results\cite{hill97,kart04, hill06}.  With the more prominent role played by AMRO experimental techniques comes a greater need to examine each assumption underpinning the theoretical models used to interpret these data.

Theoretical expressions for AMRO are most simply obtained within Boltzmann transport theory,
which is valid when the quasiparticle interlayer
transport is coherent.  The calculated magnetoresistance in strong magnetic fields is sensitive to the shape of the Fermi surface since electrons complete many cyclotron orbits before being scattered.  Using AMRO data taken over a range of $\theta_B$, $\phi_B$ and field strength, it is possible to extract multiple hopping parameters and thus build a detailed topography of the intralayer Fermi surface.  For overdoped cuprates, the results have shown good agreement with
ARPES measurements and other determinations\cite{anal07}.

In contrast to the close consideration given to the effects of complex Fermi surface shape
on AMRO, electron-scattering effects have usually been treated simply: a single
scattering rate, used as an additional fitting parameter, is assumed.  In other words, the Boltzmann theory of AMRO is treated within the relaxation-time approximation\cite{ashmerm}.

Since the net change due to scattering of the electron distribution at a given momentum depends on the distribution elsewhere, a general distribution function must be found self-consistently from the Boltzmann transport equation with the full collision functional included\cite{abrik,ziman}.  Within the relaxation-time approximation, the collision functional is assumed to be  proportional to the nonequilibrium part of the electron distribution, with a proportionality constant defined as the current relaxation rate, or inverse current lifetime.  The distribution is then easily found from the simplified transport equation.  (The relaxation-time approximation is equivalent to neglecting vertex corrections in a Kubo calculation of the conductivity\cite{rammer}.)  The electron distribution obtained this way is rarely a solution to the full Boltzmann equation, so the calculated transport coefficients are not always reliable.  Even if experimental properties can be explained using this approach, the physical meaning of the lifetime extracted from fits to data is not always clear.

It is well known that lifetimes of electrons in a given material determined from different experiments on the same sample can vary widely.   For example the scattering rate determined from zero-field transport, the transport relaxation rate $\tau_{tr}^{-1}$, is expected to be smaller than that determined from spectral or quantum oscillation measurements, the quasiparticle scattering rate $\tau^{-1}$, because small-angle scattering processes contribute only to the latter\cite{cole89,shoe84}.  This difference persists to the lowest temperatures if inhomogeneity with a large characteristic length scale is present in the sample, as may be the case for many low-dimensional metals\cite{yuds07,poly02}.

AMRO is one measurement where the physical interpretation of an extracted lifetime might not be immediately clear. For, although it is a transport measurement, the fact that electrons are bound in tight electron orbits and can be transferred from one orbit to another via small-angle scattering suggests that scattering processes which play no role in zero-field transport might be significant.  Notably, the precise meaning of lifetimes extracted from quantum oscillations in magnetotransport and from cyclotron resonance experiments is still a matter of discussion in the literature\cite{sing07, powe04, sing03, kart02, hill97, cole89,shoe84} If a lifetime determined from AMRO data using a relaxation-time approximation is to be usefully compared with those obtained elsewhere then its physical significance must be clarified\cite{anal07,sing07,abde06}.  Addressing this issue is the main purpose of this paper.

We calculate AMRO for simple model (quasi 1D and 2D) Fermi liquid metals using Boltzmann theory with the full elastic collision functional included.  Our purpose is (i) to elucidate the meaning of lifetimes extracted from AMRO by tracing their origin from the collision functional, and (ii) to determine whether new information can be extracted from AMRO by going beyond the relaxation-time approximation.   It turns out that a single effective scattering rate does not fully account for the effect of scattering on AMRO but rather, in a crude sense, the effective scattering rate relevant to AMRO depends on the angle $\theta_B$ between the field and current.  Also, we find that AMRO can provide a description of the in-layer momentum (and corresponding spatial) dependence of the scattering cross-section that is not available from other techniques.  Thus AMRO may prove useful for studying an unknown scattering mechanism.

We here briefly summarize our analysis and results before providing details.  The collision functional for a quasi-2D metal that is isotropic in the (x-y) plane is characterized with real parameters $\lambda_n$, where $n$ is an integer, defined by
\begin{equation}
\label{lambda_intro}
\lambda_{n}=\lambda_\infty\int dq_z d\phi P(|q_z|,|\phi|)(1-\mathrm{e}^{iq_zc+in\phi}).
\end{equation}
where the integral is over the (quasi-cylindrical) Fermi surface and $P(|q_z|,|\phi|)$ is the probability per unit area for scattering an electron between points on the Fermi surface separated by $q_z$,$\phi$. The scattering probability depends only on the magnitude of the change of each cylindrical momentum component. Each $\lambda_n$, having the dimensions of inverse time, is a scattering rate of potential physical significance.  The transport relaxation rate is $\tau^{-1}_{tr}=\lambda_0$ whereas the total quasiparticle scattering rate is $\tau^{-1}=\lambda_\infty$.  The $\lambda_n$ for finite $n$ contain additional information about the dominant scattering mechanism.  For example, the scale $n$ over which $\lambda_n$ approaches $\lambda_\infty$ is roughly equal to the largest length scale over which the scattering potential varies within an $x-y$ layer, measured in lattice constants (see Appendix).  This type of parametrization of the collision integral is well known, and has been used, for example, to describe the effect of small-angle impurity scattering on the Weiss oscillations in a two-dimensional electron gas.\cite{mirl98}

Our main finding is that AMRO is sensitive to $\lambda_0$ when the field is parallel to the current, to $\lambda_\infty$ when the field is perpendicular to the current, and to intermediate $\lambda_n$ at intermediate angles.  There are two immediate implications of this result.  First, it clarifies the meaning of the scattering rate extracted from AMRO fits.  Second, it demonstrates that AMRO measurements provide information about the momentum-transfer dependence of the scattering probability that may not be otherwise available.  For, as long as the scattering potential extends over more than one layer, so that $\lambda_0\neq\lambda_\infty$, it should in principle be possible to extract more than one $\lambda_n$ using AMRO (methods for isolating individual $\lambda_n$ are described below).  This would enable a partial reconstruction of the scattering cross-section via Eq. \ref{lambda_intro} and could reveal, for example, whether an unknown scatterer had atomic-length scale correlations in the layer or whether it was correlated over much larger distances.

In the next section we give the Boltzmann derivation of the interlayer conductivity for the quasi-2D and quasi-1D metals.  In Section III we give a simple physical picture of this result.  In the subsections of Section IV we analyze the conductivity in different limits, describe how individual $\lambda_n$ could in principle be extracted from AMRO data, and illustrate the expected behaviour.  In the Appendix we consider the simple example of random impurity scattering to illustrate the significance of the $n$ dependence of $\lambda_n$.

\section{Boltzmann theory of interlayer magnetoresistance with elastic, anisotropic scattering}

To calculate the resistance in a strong applied magnetic field we adopt the following form of the Boltzmann equation\cite{abrik,scho00}, which is lowest order in the strength of the electric field ${\bf E}$ but valid for an arbitrary-strength static magnetic field ${\bf B}$,
\begin{equation}
\label{boltz}
\frac{\partial g_{\bf k}}{\partial
 t}-I[g_{\bf k}]=-e{\bf E}\cdot {\bf v_k}
\end{equation}
where $g_{\bf k}$ is the non-equilibrium part of the full electron distribution function $f_{\bf k}$,
\begin{equation}
f_{\bf k}=f_0(\epsilon_{\bf k})+\bigg{[}-\frac{\partial f_0(\epsilon_{\bf k})}{\partial \epsilon_{\bf k}}\bigg{]}g_{\bf k},
\nonumber
\end{equation}
$f_0(\epsilon_{\bf k})$ is the Fermi distribution, and ${\bf v_k}=(d\epsilon_{\bf k}/d{\bf k})$.  The collision integral $I[g_{\bf k}]$ is of the form
\begin{equation}
\label{collI}
I[g_{\bf k}]=\int dS_{\bf k^{\prime}} P({\bf k},{\bf k}^{\prime}) (g_{{\bf k^\prime}}-g_{\bf k}).
\end{equation}
where $P({\bf k},{\bf k}^{\prime})$ is the probability per unit time for an electron to be scattered from ${\bf k}$ to
${\bf k^{\prime}}$, where both lie on the Fermi surface.  The integral is over the ${\bf k^\prime}$ Fermi surface with $dS_{\bf k^{\prime}}=d^2k^{\prime}/|{\bf v_{k^{\prime}}}|$.  This collision integral is appropriate for elastic scattering, which will henceforth be assumed.  Well-known approximate extensions to a given inelastic scattering mechanism could be used to predict, for example, different temperature dependencies for different $\lambda_n$ (as occurs for scattering by acoustic thermal phonons for $T<<T_D$ where $\lambda_0\approx(T/T_D)^2\lambda_\infty$)\cite{abrik}.  For electrical conductivity (as opposed to thermal conductivity) the elastic collision integral may be expected to give qualitatively correct results with the $\lambda_n$ temperature-dependent.

Under the applied magnetic field electrons move along cyclotron orbits on the Fermi surface (to lowest order in $|{\bf E}|$ the electric field can be ignored in this orbital motion) such that ${\bf k}={\bf k}(t)$ varies with the time variable $t$ according to
\begin{equation}
\label{motion}
\frac{d {\bf k}}{d t}=-e{\bf v_k}\times{\bf B},
\end{equation}
with $\hbar=1$.

To calculate the magnetoconductivity, we solve Eq. \ref{cond} to obtain the momentum ${\bf k}(t)$ for an arbitrary initial value ${\bf k}(0)$ and insert the result into Eq. \ref{boltz} to determine the distribution function $g_{\bf k}$ (a single value of $t$ is used for the momenta ${\bf k}(t)$ and ${\bf k}^{\prime}(t)$ in the collision integral).    Both ${\bf v_k}$ and $g_{\bf k}$ depend on the initial momentum ${\bf k}(0)$ and vary periodically with time.  The time-dependent current is
\begin{equation}
\label{current}
{\bf j}(t)=\frac{2e}{(2\pi)^2}\int dS_{\bf k} g_{\bf k}(t){\bf v_k}(t),
\end{equation}
where the integral over the Fermi surface is done using the initial momentum ${\bf k}(0)$ coordinates.  The frequency-dependent conductivity $\sigma(\omega)$ (for both current and electric field along $z$) is finally obtained from the time-frequency Fourier series of the $z$ component of Eq. \ref{current}.

Below we follow this method to obtain the interlayer conductivity for the quasi-2D system (Eq. \ref{cond}) and quasi-1D system (Eq. \ref{cond1D}).  These two Eqs. are the main results of this article.

\subsection{Quasi-2D metal}

 We consider a quasi-2D Fermi surface that is isotropic in the layer (the $k_x-k_y$ or $a-b$ plane) and weakly corrugated in the direction perpendicular to the layers ($k_z$ or $c$) with band energy
\begin{equation}
\label{band}
\epsilon_{\bf k}=\frac{1}{2m^*}(k_x^2+k_y^2-k_f^2)-2t_c\cos(k_zc)
\end{equation}
where $m^*$ is the effective mass and the last term is obtained from nearest-interlayer-neighbor hopping with coefficient $t_c$ and interlayer distance $c$.  It is assumed that $k_f^2/2m^*>>t_c$ where $k_f$ is the average radius of the near-cylindrical Fermi surface.

We substitute Eq. \ref{band} into Eq. \ref{motion} with the magnetic field ${\bf B}=B(\sin\theta ,0,\cos\theta )$. To lowest order in $v_z$, the $c$ axis dispersion of the Fermi surface can be ignored in the determination of the cyclotron trajectories and the collision integral so that ${\bf k}=(k_f\cos\phi ,k_f\sin\phi , k_z)$.  The $z$-component of the equation of motion Eq. \ref{motion} is
 \begin{equation}
 \label{motion2}
 \frac{\partial k_z}{\partial t}=\omega_C k_f \tan\theta_B \sin\phi
 \end{equation}
 and, dropping an interlayer hopping term under the assumption $(k_f/m^*)\sin(\phi)>>t_cc\tan\theta_B$, the intralayer components become
 \begin{equation}
 \label{motion3}
 \frac{\partial \phi}{\partial t}=\omega_C
 \end{equation}
where the cyclotron frequency is
\begin{equation}
\omega_C=(eB/m^*)\cos\theta_B.
\end{equation}
Eqs. \ref{motion2} and \ref{motion3} are solved to give
\begin{equation}
\label{vz}
v_z(t)=2t_cc\sin\bigg{[}k_zc-k_fc\tan\theta_B[\cos(\phi+\omega_Ct)-\cos\phi]\bigg{]},
\end{equation}
which can be used in the Boltzmann equation, Eq. \ref{boltz}.   The omitted interlayer hopping term is typically small, significant only for electrons moving in the direction of the magnetic field when the field is nearly
parallel to the layers.  This term gives rise to a small coherence peak in the resistivity at $\theta_B=\pi/2$ in strong magnetic field but otherwise has little effect.\cite{mose99,kart04,scho00}

The distribution function and velocity are expanded in a Fourier series over both momentum variables at $t=0$:
\begin{equation}
\label{fourier}
g_{\bf k}=\sum_{mn}g_{mn}(t) \mathrm{e}^{imk_z(0)c+in\phi(0)},
\end{equation}
\begin{equation}
\nonumber
v_z({\bf k})=\sum_{mn}u_{mn}(t)\mathrm{e}^{imk_z(0)+in\phi(0)}.
\end{equation}
To henceforth avoid such cluttered notation, we use the symbols $k_z$ and $\phi$ to refer to the initial values $k_z(0)$ and $\phi(0)$ and indicate $t$ dependence explicitly.

In this simple isotropic model, the scattering probability $P({\bf k^{\prime}},{\bf k})$ depends only on the change in intralayer momentum $|\phi^{\prime}-\phi|$.  Also, to lowest order in $v_z$ the scattering probability may be treated as a function of $|k_z^{\prime}-k_z|$.  The Fourier expansion
thus diagonalizes the collision integral, which can be written
\begin{equation}
I[g_{\bf k}]=-\sum_{mn} g_{mn}(t)\lambda_{mn}\mathrm{e}^{imk_zc+in\phi}
\end{equation}
where the collision parameters $\lambda_{mn}$, defined by
\begin{equation}
\label{lambda2}
\lambda_{mn}=\int_{-\pi/c}^{\pi/c} dq_z \int_0^{2\pi} d\phi P(|q_z|,|\phi|)(1-\mathrm{e}^{imq_zc+in\phi}).
\end{equation}
have the dimensions of inverse time.

The Boltzmann equation Eq. \ref{boltz} becomes
\begin{equation}
\label{ftboltz}
\bigg{(}\frac{\partial}{\partial t}+\lambda_{mn}\bigg{)}g_{mn}=-eu_{mn}(t)E(t),
\end{equation}
which has the solution
\begin{equation}
\label{dist}
g_{mn}(t)=-e\int_{-\infty}^t dt^{\prime} \mathrm{e}^{\lambda_{mn}(t^{\prime}-t)}u_{mn}(t^{\prime})E(t^{\prime}).
\end{equation}

The $m$ Fourier number corresponds to a label of the layers in real space, i.e. the set of coefficients $g_{m^{\prime}n}$, for all $n$ and a particular $m^{\prime}$, describes the difference in the electron distribution between layers separated by $m^{\prime}$ lattice constants along the $c$ crystal axis.  Since the band energy Eq. \ref{band} contained only
nearest-plane hopping terms (and interlayer motion
within a cyclotron orbit was excluded by dropping terms of order $v_z$) only $u_{1n}$ and $u_{-1n}$ terms are nonzero.  So, only the $m=\pm 1$ terms in
the distribution function are present and we can define single-subscript collision parameters $\lambda_n\equiv
\lambda_{1n}=\lambda_{-1n}$, as first introduced in Eq. \ref{lambda_intro}.

To carry out the $\phi$ and $t$ integrals of the nested sine functions appearing in Eqs. \ref {ftboltz} and \ref{current} it is convenient to introduce the identity e$^{iz\sin(x)}=\sum_{m=-\infty}^{+\infty} J_m(z)\mathrm{e}^{ix}$, where $J_m(z)$ is an $m$th order Bessel function of the first kind, twice into the both the
expression for $g_{\bf k}(t)$ and $v_z(t)$. (Note, in the quasi-2D and quasi-1D models, the relationship between the corresponding components of the conductivity and resistivity tensors is simply $\rho_{zz}=(\sigma_{zz})^{-1}$; we henceforth denote $\sigma_{zz}$ by $\sigma_{\perp}$.) The final result for the conductivity is
{\begin{equation}
\label{cond}
\frac{\sigma_{\perp}(\omega)}{\sigma_{\perp 0}\lambda_0}=\sum_{n,p=-\infty}^{+\infty}J^2_p(\gamma)J^2_{p-n}(\gamma)\frac{1}{\lambda_{n}}\frac{1}{1+(\frac{\omega-p\omega_C}{\lambda_{n}})^2}
\end{equation}
where the argument of the Bessel function
\begin{equation}
\gamma=k_fc\tan\theta_B,
\end{equation}
the cyclotron frequency $\omega_C$ is
\begin{equation}
\omega_C=\omega_0\cos\theta_B,
\end{equation}
where $\omega_0=(eB/m^*)$ and $\sigma_{\perp 0}$ is the zero field c-axis D.C. conductivity, given by
\begin{equation}
\label{dc}
\sigma_{\perp 0}=\frac{2e^2t_c^2ck_f}{\pi v_f}\frac{1}{\lambda_0}.
\end{equation}

\subsection{Quasi 1D metal}
To model a quasi-1D system we use the band energy
\begin{equation}
\epsilon_{\bf k}=v_f(|k_x|-k_f)-2t_b\cos(k_yb)-2t_c\cos(k_zc)
\end{equation}
where both interchain hopping parameters $t_b$ and $t_c$ are small compared to $v_fk_f$.
The Fermi surface consists of two sheets, located near $k_x=\pm k_f$, that are weakly dependent on $k_y$
and $k_z$.  Solving the Eqs. of motion, with a magnetic field ${\bf B}=B(\sin\theta,0,\cos\theta)$, we obtain
for the $c$-axis velocity on the $k_x=\pm k_f$ sheets
\begin{equation}
v^{\pm}_z(t)=2t_cc\sin\bigg{[}k_zc\mp k_fc\tan\theta[\cos(k_yb\pm\omega_Ct)-\cos k_yb]\bigg{]}.
\end{equation}

The $c$-axis conductivity is calculated in much the same way as for the quasi 2D system.  A slight complication results from the need to consider both intrasheet and intersheet scattering processes.  If we write $g^\pm_{\bf k}$ for the distribution on the $k_x=\pm k_f$ sheet then intersheet scattering in the collision integral couples $g^+_{\bf k}$ with $g^-_{\bf k}$.  However, Fourier expansions of the sum $g^s_{\bf k}=g^+_{\bf k}+g^-_{\bf k}$ and difference $g^d_{\bf k}=g_{\bf k}^+-g_{\bf k}^-$ diagonalize their respective collision integrals.  The Fourier components of the distributions are determined from
\begin{equation}
\label{ftboltz1D}
\bigg{(}\frac{\partial}{\partial t}+\lambda^{\nu}_{mn}\bigg{)}g^{\nu}_{mn}=-eu^{\nu}_{mn}(t)E(t),
\end{equation}
where $\nu=s,d$ and $u^s_{mn}$ and $u^d_{mn}$ are the Fourier series for $v_z^+({\bf k})+v_z^-({\bf k})$ and $v_z^+({\bf k})-v_z^-({\bf k})$, respectively.  The collision parameters are given by
\begin{equation}
\label{lambdas}
\lambda^s_{mn}=\int dq_ydq_z P(|q_y|,|q_z|)(1-\mathrm{e}^{imq_zc+inq_yb})
\end{equation}
and
\begin{equation}
\label{lambdad}
\lambda^d_{mn}=\lambda^s_{mn}+2\int dq_ydq_z P_1(|q_z|,|q_y^|)\mathrm{e}^{imq_zc+inq_yb},
\end{equation}
where $P({\bf k})=P_0({\bf k})+P_1({\bf k})$ is the total scattering probability per unit time, $P_0({\bf k})$ the intrasheet and $P_1({\bf k})$ the intersheet component.  Just like for the quasi 2D case, only $m=\pm 1$ terms are important so we define single-subscript collision parameters $\lambda_n^\nu\equiv\lambda^\nu_{1n}=\lambda^\nu_{-1n}$.

The remainder of the calculation is identical to the 2D case.  The final result is
\begin{equation}
\label{cond1D}
\frac{\sigma_{\perp}(\omega)}{\sigma_{\perp 0}\lambda_0}=\frac{1}{2}\sum_{np\;\nu}J^2_p(\gamma)J^2_{p-n}(\gamma)\frac{1}{\lambda^\nu_n}\frac{\eta_\nu}{1+(\frac{\omega-p\omega_C}{\lambda^\nu_n})^2}
\end{equation}
where $\eta_s=1+(-1)^n$ and $\eta_d=1-(-1)^n$.  The argument of the Bessel function is $\gamma=2(t_bc/v_f)\tan\theta_B$ and the prefactor
\begin{equation}
\sigma_{\perp 0}\lambda_0=\frac{2e^2t_c^2c}{v_fb\pi}.
\end{equation}
Note that the presence of the factor $t_bc/v_f$, which will be smaller than one for quasi-1D systems, in the argument of the Bessel function suggests that observable AMRO will be restricted to higher values of $\theta_B$ than in the quasi-2D case.

If intersheet scattering is ignored, so $\lambda_n^s=\lambda_n^d$, then Eq. \ref{cond1D} reduces to the form of Eq. \ref{cond}.

\subsection{Relaxation-time approximation}

If the collision integral is treated using the relaxation time approximation then the $n$ dependence of $\lambda_n$ is ignored and a single transport relaxation rate, say $\tau^{-1}_{B}$ is assumed.  If we set all $\lambda_n$ equal to $\tau_B^{-1}$ in Eq. \ref{cond} or \ref{cond1D} (in the latter, no distinction is made between intrasheet and intersheet scattering) then, using the Bessel function identity $\sum_{m=-\infty}^{+\infty}J_m^2(x)=1$, both Eqs. reduce to
\begin{equation}
\label{rtcond}
\frac{\sigma_\perp(\omega)}{\sigma_{\perp 0}}=\sum_{p=-\infty}^\infty J^2_p(\gamma)\frac{1}{1+(\omega-p\omega_C)^2\tau_B^2},
\end{equation}
which has been found previously by Moses and McKenzie\cite{mose99}.

\subsection{Contrast with the intralayer conductivity}

The intralayer conductivity component $\sigma_{\parallel}\equiv\sigma_{xx}$ may easily be calculated within the same simple models.  Taking the electric field to lie purely along the $\hat{{\bf x}}$-axis, and repeating the above calculation for the quasi-2D metal, we find:
\begin{equation}
\label{drude}
\frac{\sigma_{\parallel}(\omega)}{\sigma_{\parallel 0}\lambda_{01}}=\frac{\lambda_{01}}{\lambda_{01}^2+(\omega-\omega_C)^2}
\end{equation}
where $\sigma_{\parallel 0}\lambda_{01}=4\pi e^2Ek_f/(mc)$.  In this expression, there are no angular-dependent oscillations and the only relevant collision parameter is the intralayer transport relaxation rate $\lambda_{01}$.  All the novel effects in the inter-layer transport, Eq. \ref{cond}, arise from the non-trivial $k_z$-dispersion, and are absent from the intralayer transport coefficients in this model.

For the quasi 1D system, the $yy$ component of the conductivity is of the same form as Eq. \ref{cond1D} (with the $b$ and $c$ axes interchanged).  For the ${xx}$ component, since the velocity $v^{\pm}_x=\pm k_f/m^*$, the right side of Eq. \ref{ftboltz1D} is equal to $0$ for $\nu=s$ and to $-2eEk_f/m^*$ for $\nu=d$.  The result for the quasi-1D conductivity $\sigma_{1D\parallel}(\omega)$ is thus
\begin{equation}
\label{drude1D}
\frac{\sigma_{\parallel}(\omega)}{\sigma_{\parallel 0}\lambda^d_{00}}=\frac{\lambda^d_{00}}{(\lambda^d_{00})^2+\omega^2}.
\end{equation}
where the prefactor is $\sigma_{\parallel 0}\lambda^d_{00}=e^2(k_f/m^*)^2/(\pi bc)$.

The collision parameters that enter the interlayer conductivity $\lambda_n\equiv\lambda_{1n}$ describe the relaxation of a difference in current density on adjacent layers.  They do not include any of the $\lambda_{0n}$ parameters, which describe the relaxation of current density variation within a single layer and of which the intralayer transport relaxation rate $\lambda_{01}$ is one member.  Nevertheless the fact that $\lambda_{1\infty}=\lambda_{0\infty}=\lambda_{m \infty}$ implies that the interlayer conductivity is sensitive to the total quasiparticle relaxation rate in a tilted field (indeed, as discussed below, the total quasiparticle relaxation rate is the relevant quantity when the magnetic field is parallel to the layers).

\section{Physical picture of scattering and AMRO}

\begin{figure}

\begin{center}
\includegraphics[width=2.0 in, height=2.5 in]{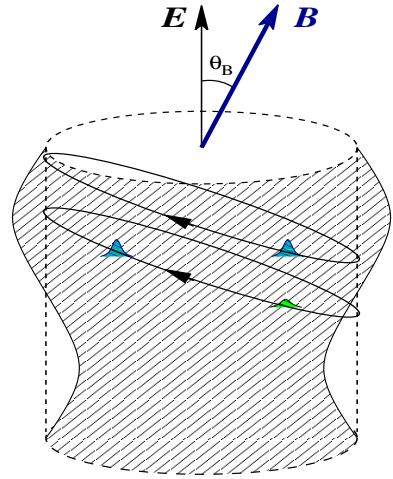}
\end{center}
\caption{(Color online.) \label{sketch} Interlayer current relaxation of a quasi 2D metal in a tilted magnetic field. The Fermi surface is shown as the dashed cylinder and the electron distribution in an interlayer (i.e. $\parallel k_z$) electric field as the hatched region.  Electrons undergo cyclotron motion in the magnetic field, tilted by $\theta_B$ from the current direction.  The two upper (blue) bundles of electrons are initially at the same $k_z$ and have the same density.  For $\theta_B\neq 0$ their orbits separate the bundles in $k_z$, which allows scattering $\perp k_z$ to relax the current of each.  Bundles on the right (blue and green), initially displaced in $k_z$ with different densities, orbit in phase so scattering $\parallel k_z$ always exchanges electrons between them.}
\end{figure}

Before proceeeding to analyze Eq. \ref{cond} in more detail, it is worth discussing its qualitative features (only the quasi-2D system for $\omega=0$ will be considered).  The oscillatory behaviour of the magnetoresistance, which is captured by Eq. \ref{rtcond}, has been discussed elsewhere\cite{mose99}.  Here we focus on the effect of scattering and, in particular, on the difference between the result obtained using the full collision functional Eq. \ref{cond} and that found using the relaxation time approximation Eq. \ref{rtcond}.  The difference results from the appearance of multiple collision parameters $\lambda_n$ in Eq. \ref{cond}.  For this discussion it is helpful to recall that small-angle scattering within a $k_x-k_y$ plane contributes to $\lambda_n$ for large $n$ but has no effect on $\lambda_0$ (i.e. the integrand in Eq. \ref{lambda_intro} vanishes when $n=0$ and $q_z$=0).

The factor $J_p^2(\gamma)$ in Eq. \ref{cond} originates from the Fourier expansion in $\phi$ of the interlayer current distribution $g(k_z,\phi,t)v_z(k_z,\phi,t)$ at the zero of time while $J^2_{p-n}(\gamma)$ comes from the current distribution at a different time.  So Eq. \ref{cond} has the form of a current-current correlation function expected from the Kubo formula\cite{mahan}.  The ``incoming'' current (that associated with $p$) and ``outgoing'' current (associated with $p-n$) are coupled by a non-trivial vertex involving collision parameters $\lambda_n$.  In the relaxation time approximation, the non-trivial vertex factors are ignored so the currents decouple and the expression reduces to Eq. \ref{rtcond}.

A Bessel function $J_p(x)$ becomes small once its order $p$ exceeds its argument $x$, so the terms $p$ that contribute to Eq. \ref{cond} are those for which $p\lessapprox k_fc\tan\theta_B$.  This means that the incoming current distribution varies more rapidly around the cylindrical Fermi surface with increasing $\theta_B$.  Since $|p-n| \lessapprox k_fc\tan\theta$, the outgoing current distribution also varies more rapidly, which enables collision parameters $\lambda_n$ for larger $n$ to become involved. Thus AMRO becomes more sensitive to small-angle scattering in the $k_x-k_y$ plane as $\theta_B$ is increased.  This behavior can be better appreciated using the following simple picture.

In the absence of a magnetic field the conductivity is Eq. \ref{dc} and the nonequilibrium part of the electron distribution is
\begin{equation}
\label{gsimp}
g_{\bf k}=eEv_z\lambda_0^{-1},
\end{equation}
which is shown as the hatched region in Fig. \ref{sketch}.  It is independent of $\phi$, so a Fourier expansion of the current distribution in $\phi$ has only the $p=0$ term.  Scattering perpendicular to $\hat{\bf k}_z$ connects points of equal current distribution, with no net effect, so only $\lambda_0$ (no other $\lambda_n$) appears.

Now suppose we start from the $B=0$ distribution Eq. \ref{gsimp} and, at $t=0$, turn on a strong magnetic field at an angle $\theta_B$ from the interlayer electric field.  Electrons make cyclotron orbits as shown in Fig. \ref{sketch}.  Two bundles of electrons that initially had the same $k_z$ (and thus the same density $g_{\bf k}$) but different $\phi$ will move apart in $k_z$ as they follow their respective orbits.  In the same way, bundles of unequal densities are brought together onto a given $k_z=const$ plane.  The current distribution at a given $k_z$ thus becomes dependent on $\phi$ at times $t>0$. If $\theta_B$ is large so the orbits are elongated, this variation with $\phi$ is rapid because bundles of widely different density are being brought together.  This is the reason why increasingly large $p$ contribute in Eq. \ref{cond} as $\theta_B$ is increased.

If the current distribution varies rapidly as a function of $\phi$ then small-angle scattering within a $k_z=const$ plane can effectively relax the current.  Inspecting Eq. \ref{cond} we see that this is what happens near $\theta_B=\pi/2$.  For, as $\theta_B$ increases, terms in the sum for which $p$ and $p-n$ differ substantially appear.  The incoming and outgoing current distributions then differ at points narrowly separated in $\phi$, which allows $\lambda_n$ for large $n$, i.e. small-angle scattering along a $k_x-k_y$ plane, to affect the conductivity.

 While the magnetoresistance becomes more sensitive to small angle scattering perpendicular to $\hat{\bf k}_z$ as $\theta_B$ increases, there is no change associated with scattering parallel to $\hat{\bf k}_z$.  This may be understood by following the evolution of two bundles of electrons initially at the same $\phi$ but different $k_z$. When we turn on the field at $t=0$, the bundles follow orbits displaced in $k_z$ while their phases $\phi$ remain equal.  Scattering along $k_z$ transfers an electron from one bundle to the other just as it did at $t=0$.  This is true regardless of the value of $\theta_B$.  So scattering with momentum transfer parallel to $k_z$ is equally effective at all $\theta_B$.  (Of course, this qualitative anisotropy in the effect of scattering $\perp$ and $\parallel$ to $k_z$ is a consequence of our dropping higher order terms in $v_z$ for the quasi-2D system.)

These simple considerations are sufficient to understand the qualitative effect of scattering in AMRO for the quasi-2D system.  The interlayer magnetoresistance becomes increasingly sensitive to small-angle scattering in the momentum plane perpendicular to the current as the field angle $\theta_B$ is increased.  This is why AMRO might prove a useful probe of the intralayer properties of scatterers.  This physics is missed entirely in the relaxation time approximation, which treats all scattering processes as equivalent.

\section{Analysis of AMRO and the extraction of collision parameters}

 The accuracy of the relaxation time approximation in AMRO will depend on the nature of the dominant scattering mechanism.  For scattering by atomic-scale defects, $\lambda_0$ and $\lambda_\infty$ will differ by at most a factor of order unity.  If the dominant scatterer is of electronic origin then, for the anisotropic systems under consideration, the scattering potential will be expected to have only short-range interlayer correlations and, once again, $\lambda_0\approx\lambda_\infty$.  Only when the scattering is due to long-wavelength phonons or other inhomogeneities  extending over many layers will there be strong $n$-dependence in the collision parameters $\lambda_n$.

 Thus the relaxation-time approximation should give qualitatively correct results for interlayer transport in low-dimensional systems for most cases of interest.  But given the quantitative accuracy with which the Fermi surface and other properties have been measured using AMRO, a more precise description of the effect of scattering appears prudent.  Also, as long as there is an observable difference between $\lambda_0$ and $\lambda_\infty$,  we can take advantage of the sensitivity of Eqs. \ref{cond} and \ref{cond1D} to additional collision parameters in order to extract information about the scatterer.

 For the remainder of this section we analyze the behaviour of Eqs. \ref{cond} and \ref{cond1D}.  We are mainly interested in how $\lambda_n$ might be extracted from experiment in various limiting cases.

\subsection{Field along the layers}

If the magnetic field is directed along the c-axis then only the zeroeth order Bessel functions contribute to the sum and Eqs. \ref{cond} and \ref{cond1D} reduce to the zero-field conductivity $\sigma(0)=\sigma_0$, i.e. it is sensitive to $\lambda_0$ alone.  As $\theta_B$ approaches $\pi/2$ a large number of terms contribute and the expression is dominated by large $n$ and $p$.  Here, $\lambda_n$ can be replaced by $\lambda_{\infty}$ and the $n$ sum easily done. Both  the $p$th order Bessel function and the Lorentzian are capable of cutting off the sum over $p$ and which one actually imposes the cutoff depends on the size of $\Omega_C\lambda_{\infty}^{-1}$ where $\Omega_C=k_fc(eB/m^*)$.  The $\pi/2$ limit of Eq. \ref{cond} is, for $\omega<<\Omega_C,\lambda_\infty$
\begin{equation}
\label{pi2}
\frac{\sigma_\perp(\omega=0)}{\sigma_{\perp 0}\lambda_0}=\frac{1}{\lambda_{\infty}}\frac{1}{\sqrt{1+(\Omega_C/\lambda_{\infty})^2}}\;\;\;\;\;\;\;\;\;\theta_B=\pi/2
\end{equation}
(This result has previously been obtained within the relaxation
time approximation by Schofield\cite{scho00}).  The 1D result is of the same form with $\Omega_C=2(eB/m^*)t_b/(\hbar v_f)$ since, as seen from Eqs. \ref{lambdas} and \ref{lambdad}, $\lambda_\infty^s=\lambda_\infty^d$.

In the $\Omega_C/\lambda_{\infty}>>1$ limit of Eq. \ref{pi2}, the conductivity is inversely proportional to field and independent of scattering, i.e of $\lambda_\infty$.  In the opposite limit, $\Omega_C/\lambda_\infty<<1$ the conductivity is then proportional to $1/\lambda_\infty$, and independent of the field.

Comparing the $\theta_B=0$ and $\theta_B=\pi/2$ limits of Eq. \ref{cond} we see that the $c$-axis magnetoresistance is sensitive to either the transport relaxation rate $\lambda_0$ or the quasiparticle scattering rate $\lambda_{\infty}$ depending on whether  the field-angle $\theta_B$ is perpendicular or parallel to the metallic layers.  Information about $\lambda_n$ for finite $n$ is available at intermediate angles.

\subsection{Strong-field limit $(\omega_C>>\lambda_n)$}

AMRO is seen when $\omega_C>>\lambda_0$ at small $\theta_B$.  In this case, the sum over $p$ in Eq. \ref{cond} will be dominated by the $p=0$ term except very close to the conductivity minima (where the zeroeth order Bessel function vanishes) and close to $\pi/2$. Everywhere else, the conductivity will be given by
\begin{equation}
\label{approxcond}
\frac{\sigma_{\perp}(\omega=0)}{\sigma_{\perp 0}\lambda_0}\approx\frac{J_0^2(\gamma)}{\lambda_0}\bigg{(}J_0^2(\gamma)+\frac{J_1^2(\gamma)}{\lambda_1}+\frac{J_2^2(\gamma)}{\lambda_2}+....\bigg{)}.
\end{equation}
The corresponding expression for the 1D metal is
\begin{equation}
\label{approxcond1D}
\frac{\sigma_{\perp}(\omega=0)}{\sigma_{\perp 0}\lambda_0}\approx\frac{J_0^2(\gamma)}{\lambda^s_0}\bigg{(}J_0^2(\gamma)+\frac{J_1^2(\gamma)}{\lambda^d_1}+\frac{J_2^2(\gamma)}{\lambda^s_2}+....\bigg{)},
\end{equation}
with the $\lambda^s_n$ parameters associated with the even-order and $\lambda^d_n$ parameters with the odd-order Bessel functions.

\subsection{Weak field limit $(\omega_C<<\lambda_n)$}

In the opposite limit, in which $\omega_C<<\lambda_n$ for all $n$, the $\theta_B$ dependence of
the conductivity is much weaker.  The second term in the denominator will be negligible for all $p$ and
Eq. \ref{cond} becomes
\begin{equation}
\frac{\sigma_{\perp}(\omega=0)}{\sigma_{\perp 0}\lambda_0}=\sum_{n,p}J^2_p(\gamma)J_n^2(\gamma)\lambda_{p+n}^{-1}.
\end{equation}
This is a weighted average of $\lambda^{-1}_n$ for $n<n_{\mathrm{max}}\approx
k_fc\tan(\theta_B)$.
It suggests that AMRO fits using a single scattering parameter might
be improved somewhat by allowing the scattering rate to depend on field-angle $\theta_B$, although this is clearly a
crude treatment.

For the quasi-1D case, the expression in this limit is very similar
\begin{equation}
\frac{\sigma_{\perp}(\omega=0)}{\sigma_{\perp 0}\lambda_0}=\frac{1}{2}\sum_{n,p}J^2_p(\gamma)J_n^2(\gamma)(\eta^s(\lambda^s_{p+n})^{-1}+\eta^d(\lambda^d_{p+n})^{-1}).
\end{equation}

\subsection{Finite frequency conductivity}
\label{secw}

 The $\omega$-dependent conductivity provides additional routes to obtain the collision
 parameters.  For a given frequency $\omega$, the conductivity will have peaks as a function
 of magnetic field strength when $\omega=p^{\prime}\omega_C$ for some integer $p^{\prime}$.  So if the field strength $B$
 is close to $B_\omega/(p^{\prime}\cos\theta_B)$ where $(eB_\omega/m^*)=\omega$, the
 conductivity will be
 \begin{equation}                                                                                \frac{\sigma_{\perp}(\omega)}{\sigma_{\perp 0}\lambda_0}=\frac{J_{p^{\prime}}^2(\gamma)}{p^{\prime}\cos\theta_B}\sum_n
 J_{p^{\prime}-n}^2(\gamma)L_{p^{\prime}n}(B)+...
 \end{equation}
 where the additional terms are smoothly field-dependent and
 \begin{equation}
 L_{p^{\prime}n}(B)=\frac{(\lambda_n/p^{\prime}\cos\theta_B)}{(\frac{\lambda_n}{p^{\prime}\cos\theta_B})^2+(\frac{eB}{m^*}-\frac{eB_\omega}{p^{\prime}m^*\cos\theta_B})^2}.
 \end{equation}
 If $\theta_B$ is reasonably small
  then the zeroeth order Bessel function in the sum over $n$ will dominate, so $n=p^{\prime}$ and the field
  dependence is a Lorentzian with a peak located at $B=B_\omega/(p^{\prime}\cos\theta_B)$ having width equal to $\lambda_{p^{\prime}}/(p^{\prime}\cos\theta_B)$.  Thus field scans
  of the conductivity measured at finite frequency and reasonably low $\theta_B$ could be used to extract individual $\lambda_n$ for small $n$.  At higher $\theta_B$ the expression
  would no longer be Lorentzian, but rather a weighted sum of Lorentzians with different widths $\lambda_n$ ($n$
 within roughly $k_fc\tan\theta_B$ of $p^{\prime}$ contribute to this weighted sum).

For the 1D metal this discussion also applies with the only difference that the width of the corresponding Lorentzian is
either $\lambda^s/(p^{\prime}\cos\theta_B)$ or $\lambda^d/(p^{\prime}\cos\theta_B)$ depending on whether $p^{\prime}$ is odd or even, respectively.

\subsection{Example: Gaussian scattering probability}

 To illustrate the behaviour of Eqs. \ref{cond} and Eq. \ref{cond1D} with a simple example we
 consider a Gaussian scattering probability with width in $q_zc$ and
 $\phi$ equal to $\sqrt{2}\Delta_z$ and $\sqrt{2}\Delta_{\phi}$ respectively.  That is, we take the scattering probability
 $P(|q_z|,|\phi|)\propto \exp[-(q_zc/2\Delta_z)^2]\exp[-(\phi/2\Delta_\phi)^2]$ so that
 \begin{equation}
 \lambda_n=\lambda_\infty(1-\mathrm{e}^{-n^2\Delta_{\phi}^2-\Delta_z^2})
 \end{equation}
 and $\lambda_0=\lambda_{\infty}(1-\mathrm{e}^{-\Delta^2_z})$.  This form is used for simplicity, but captures the qualitative characteristics of a scattering potential with two spatial length scales, one within the layers $(\Delta_\phi k_f)^{-1}$ and one perpendicular to the layers $c\Delta_z^{-1}$, for which the scattering probability is peaked at zero-momentum-transfer.  (This is illustrated in the Appendix, in which a familiar model for a random impurity potential is considered and the collision parameters determined to have the same qualitative properties as the Gaussian model used in this section.)  A plot of the collision parameter $\lambda_n$ versus its index $n$ is shown, for parameter values $\Delta_z$, $\Delta_{\phi}$ used below, in Fig. \ref{lambfig}.

\begin{figure}
\begin{center}
\includegraphics[width=3 in, height=2.5 in]{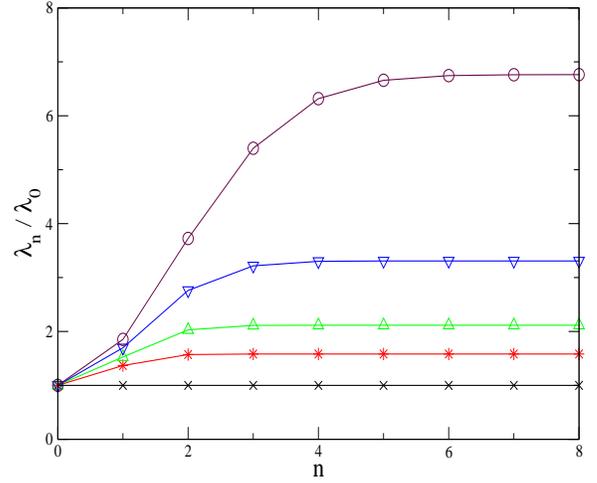}
\end{center}
\caption{\label{lambfig} (Color online.) Collision parameters $\lambda_n$ for scattering that favors small-angle scattering.  Plotted is the $n$ dependence of the collision parameters $\lambda_n$ for the Gaussian scattering potential discussed in the text.  From bottom-to-top, the curves increasingly favor small angle scattering and correspond to $\Delta_z=\Delta_\phi=\infty,1.0,0.8,0.6,0.4$, respectively.  Recall that the scale, in $n$, over which $\lambda_n$ approaches $\lambda_\infty$ corresponds to the real-space length scale of the scattering potential within the quasi-2D layer, measured in lattice constants (i.e. it corresponds to $\Delta_{\phi}^{-1}$).  The overall magnitude of the increase $\lambda_\infty/\lambda_0$ depends on the length scale perpendicular to the layers (i.e. on $\Delta_z^{-1}$).}
\end{figure}

\subsubsection{Small-angle scattering in a 2D system}

 In Fig. \ref{singtau} we illustrate the difference between the behavior of AMRO predicted by Eq. \ref{cond}, and that obtained within the relaxation time approximation Eq. \ref{rtcond}.  We choose parameters that describe a situation in which small-angle scattering is dominant: $\Delta_z=\Delta_\phi=0.3$ so that $\lambda_\infty/\lambda_0\approx 10$ in order to best illustrate qualitative behavior.  Also, we use $k_fc=3$ and $\Omega_C/\lambda_0=15$.  The curve clearly shows that the relevant scattering rate in AMRO makes a gradual transition from $\lambda_0$ to $\lambda_{\infty}$ with increasing $\theta_B$.

\begin{figure}
\begin{center}
\includegraphics[width=3 in, height=3.0 in]{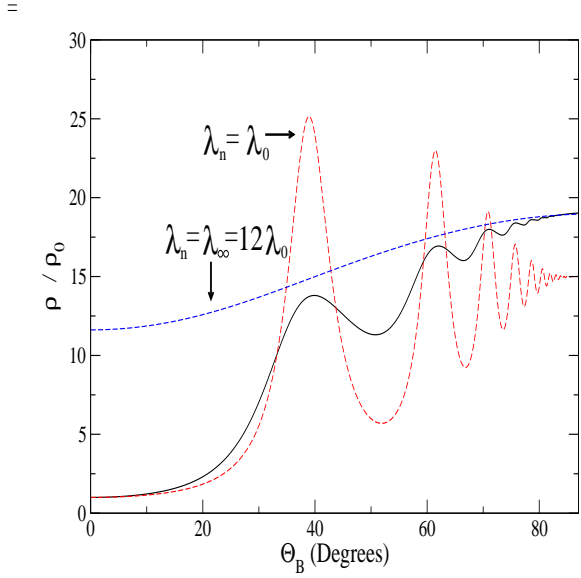}
\end{center}
\caption{\label{singtau} (Color online.) Going beyond the relaxation-time relaxation has a significant effect on AMRO when small-angle scattering is dominant.  The interlayer resistivity, obtained from Eq. \ref{cond}, of a quasi 2D metal as a function of the angle $\theta_B$ between the magnetic field and the current is shown as the solid line.  Both dashed lines are obtained using
the relaxation time approximation, i.e. by replacing the full collision functional with a single scattering rate equal to either the transport relaxation rate $\lambda_0$ or the total quasiparticle
scattering rate $\lambda_\infty$.  The parameters have been chosen such that $\lambda_\infty/\lambda_0 \approx 10$, which corresponds to dominant small angle scattering, and $\omega_C/\lambda_0=15$ at $\theta_B=0$.}
\end{figure}

\begin{figure}
\begin{center}
\includegraphics[width=3 in, height=2.5 in]{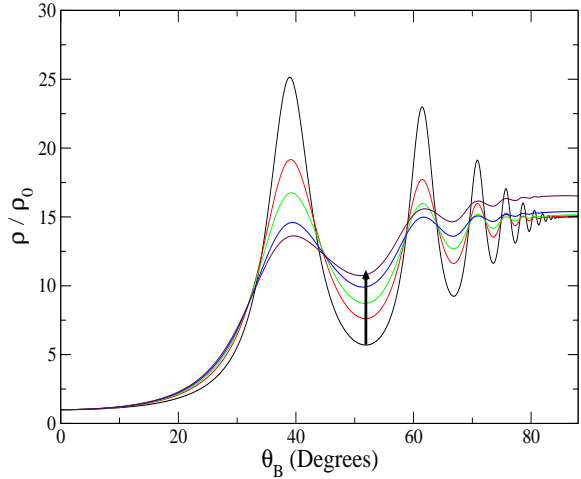}
\end{center}
\caption{\label{dccond} (Color online.) Suppression of AMRO by the removal of large-angle scattering processes.  The interlayer resistivity of a quasi-2D system is plotted as a function of magnetic field orientation for varying scattering parameters with small-angle scattering favored.  Following the arrow, the curves are for scattering that increasingly favors small-angle processes, using the same parameters as in Fig. \ref{lambfig}.  The current relaxation rate $\lambda_0$ is the same for all curves while the total quasiparticle scattering rate $\lambda_\infty$ increases by a factor of seven from top to bottom.}
\end{figure}

In Fig. \ref{dccond} we plot the resistivity for various values of the interlayer and intralayer parameters $\Delta_z$ and $\Delta_{\phi}$, taking $k_fc=3$ and $\Omega_C/\lambda_0=15$ for all the curves, and use $\Delta_z=\Delta_\phi=\infty,1,0.8,0.6,0.4$ from the bottom curve to the top along the arrow shown.  Note that the transport relaxation rate $\lambda_0/\Omega_C=1/15$ is the same for all curves but the total quasiparticle scattering rate $\lambda_\infty/\Omega_C$ increases from $1/15$ to nearly $1/2$ with ascending curves.  The resistivity at $\pi/2$ changes little, as expected in the strong-field regime.  The $n$-dependence of the collision parameters has the effect of increasingly widening and suppressing the AMRO as $\theta_B$ is increased.

\subsubsection{Large-angle scattering in quasi-1D systems}
If scattering by a spin or charge density wave with a finite ordering wavevector is important, then the scattering probability could be peaked about a particular large angle.  For the quasi-2D system, the isotropic model used here is not applicable to this case (such scattering only affects electrons near `hot spots' so the system is necessarily anisotropic).  However, we may look at the case of a quasi-1D system with a scattering mechanism that strongly favors intersheet scattering.
\begin{figure}

\begin{center}
\includegraphics[width=3 in, height=2.5 in]{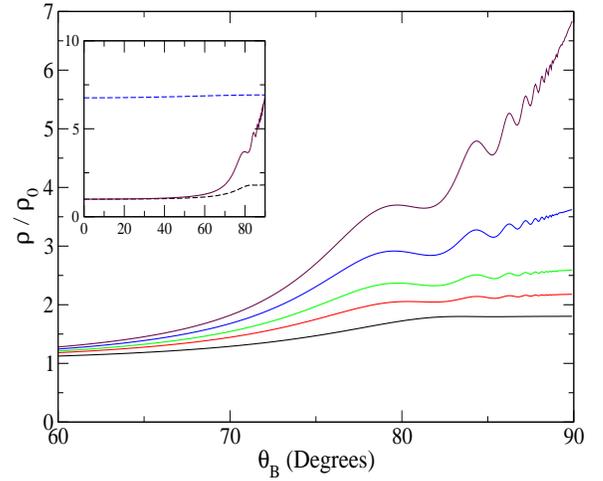}
\end{center}
\caption{\label{dccond1D} (Color online.) High-sensitivity of AMRO to momentum-dependent scattering in a quasi-1D system. Main panel: The interchain resistivity in a quasi-1D system is plotted as a function of magnetic field angle for varying scattering parameters with large-angle (intersheet) scattering favored.  From bottom to top, the curves are for scattering that increasingly favors direct scattering from one quasi 1D Fermi sheet to the other (without momentum change along the sheet). The parameters $\Delta_z$, $\Delta_\phi$ are the same as in Fig. \ref{dccond} and we have used $2t_bc/v_f=0.3$.  Inset: The solid curve is the $\Delta_z=\Delta_\phi=0.4$ curve from the main panel, plotted over the entire range of $\theta_B$, along with the relaxation-time results for $\lambda^s_n=\lambda^d_n=\lambda_0$ (lower dashed curve) and $\lambda^s_n=\lambda^d_n=\lambda_\infty$ (upper dashed curve).}
\end{figure}

 If we ignore scattering within a single Fermi surface sheet and take the intersheet scattering probability to have the Gaussian form above then we find $\lambda^s_n=\lambda_{\infty}(1-\mathrm{e}^{-n^2\Delta_{\phi}^2-\Delta_z^2})$ and $\lambda^d_n=\lambda_{\infty}(1+\mathrm{e}^{-n^2\Delta_{\phi}^2-\Delta_z^2})$.  While the former may become small for small $n$, the latter never differ significantly from $\lambda_{\infty}$.  This is because direct intersheet scattering, with no change in momentum along the sheet, does not change the sum of the electron densities of the two sheets but is effective at relaxing any difference between them.

 We show, in the main panel of Fig. \ref{dccond1D} the resulting magnetoresistance for the same scattering parameters as used in Fig. \ref{dccond}.  Here, we take the parameter $2t_bc/(\hbar v_f)=0.3$, which reflects the fact that the interchain hopping parameter $t_b$ is small compared to the intrachain Fermi energy $\hbar v_fk_f$.  This small parameter, which occurs in the argument of the Bessel function, restricts AMRO to large $\theta_B$.  For this reason, only $\theta_B$ between 60$^0$ and 90$^0$ are plotted in the main panel.  The factor $\omega_c/\lambda_0=5$ at $\theta_B=0$ as for the quasi-2D case above.

 The resistivity at $\theta_B=90^0$ increases significantly as the parameters $\Delta_z$, $\Delta_\phi$ are decreased.  This dependence occurs because the small factor $2t_bc/(\hbar v_f)$ ensures that the second term within the square root of Eq. \ref{pi2} is not dominant as it is for the quasi-2D case.  Because of this effect, which magnifies the field-angle dependence at large $\theta_B$, AMRO are more prominent for curves with small $\Delta_z,\Delta_\phi$ (i.e. for curves in which direct intersheet scattering, without momentum transfer along the sheet, is dominant).  In contrast, for the $\Delta_z=\Delta_\phi=\infty$ curve, AMRO are barely perceptible even though the system is in the strong-field limit at $\theta_B=0$.

 In the inset, the $\Delta_z=\Delta_\phi=0.4$ curve is replotted over the full $\theta_B$ range as the solid curve.  The lower and upper dashed curves show the result obtained using the relaxation-time approximation with $\tau_B^{-1}=\lambda_0$ and $\tau_B^{-1}=\lambda_\infty$, respectively.

\subsubsection{Finite-frequency conductivity}

To illustrate the qualitative frequency dependence of the conductivity Eq. \ref{cond} (or, equivalently, the magnetic field-strength dependence of the conductivity at a given finite frequency) we show, in Fig. \ref{qualcomp}, a plot of $\sigma(\omega)$ versus field $B$ at a very large frequency $\omega=24\lambda_0$ and at an angle $\theta_B=30^0$.  As above, we define a field $B_\omega$ by $\omega=eB_\omega/m^*$.  The plot is for the 2D system with Gaussian collision parameters: $\Delta_z=\Delta_\phi=1$, which corresponds to relatively weak $n$-dependence in $\lambda_n$ and gives $\lambda_1/\lambda_0=1.36$ and $\lambda_2/\lambda_0=1.55$.

The plot of $\sigma(\omega)$ has peaks located whenever the frequency $\omega$ is an integer multiple of the cyclotron frequency $\omega_C=eB\cos\theta_B/m^*$, i.e. peaks at $B\cos\theta/B_\omega=1,1/2,1/3,.. \mathrm{etc}.$  The height of these peaks drops off rapidly when $\theta_B$ is small since the amplitude of the the $p$th peak is proportional to a factor $J^2_p(k_fc\tan\theta_B)$.  So, when the argument of the Bessel function is of order unity, only the first few peaks are present.  The peaks are approximately Lorentzian with the width of the peak occurring at $\omega=p\omega_C$ being $\lambda_p/p\cos\theta_B$.

To see the evolution of $\sigma(\omega)$ with angle we show, in Fig. \ref{accond} plots of $\sigma(\omega)$ over the same field range at two frequencies: $\omega=6\lambda_0$ in the left panel and $\omega=12\omega_0$ in the right panel, for various angles: $\theta_B=5,15,30^0$.  The collision parameters are the same as in Fig. \ref{qualcomp}.
\begin{figure}
\begin{center}
\includegraphics[width=3 in, height=2.5 in]{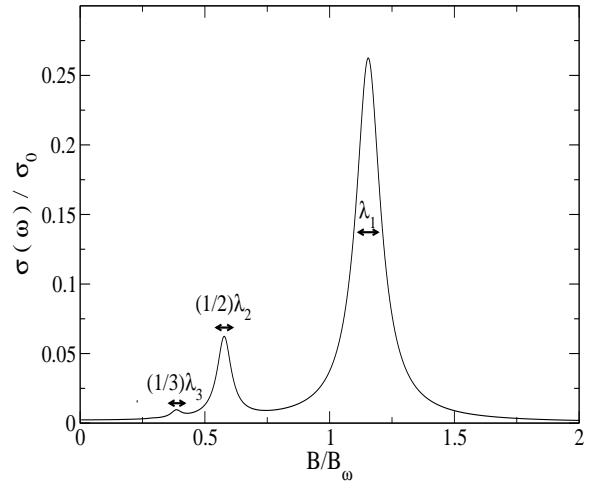}
\end{center}
\caption{\label{qualcomp} The frequency-width of resonance peaks in interlayer resistivity are determined by corresponding Fourier component of the angle-dependent scattering probability.  Plotted is the interlayer conductivity $\sigma(\omega)$, from Eq. \ref{cond}, at a frequency $\omega=24\lambda_0$ versus magnetic field strength $B$.  The conductivity has peaks occurring when the $\omega$ is an integer multiple of the cyclotron frequency $\omega_C=eB/(m\cos\theta_B)$.  We show the plot for field angle $\theta_B=30^0$ and have defined a $B_\omega$ by $\omega=eB_\omega/m$.  The peaks at $\omega=\omega_C$,$2\omega_C$ and $3\omega_C$ are visible.  Each peak is approximately Lorentzian, with a width related to the corresponding collision parameter: i.e., the respective widths of the peaks shown are $\lambda_1/\cos\theta_B, \lambda_2/2\cos\theta_B$, and $\lambda_3/3\cos\theta_B$.  }
\end{figure}

\begin{figure}

\begin{center}
\includegraphics[width=3 in, height=2.5 in]{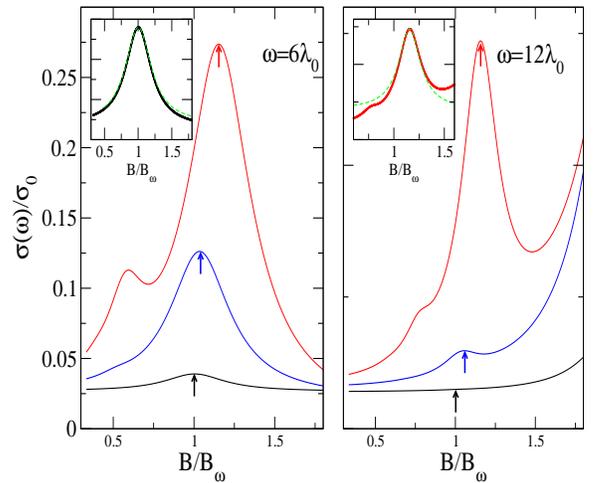}
\end{center}
\caption{\label{accond} (Color online.) The interlayer conductivity $\sigma(\omega)$ as a function of field $B$.  The frequency in the left panel is half that in the right, and the curves are for different field angles $\theta_B$: from bottom to top $\theta_B=5^0,15^0,30^0$.  The marked peaks in the left panel occur where $\omega=\omega_C$ with $\omega_C=(eB/m)\cos\theta_B$, and those in the right panel occur when $\omega=2\omega_C$.  In the left inset a zoom of the $\theta_B=5^0$ curve near $\omega=\omega_C$ is compared to a Lorentzian (dashed curve) with width $\lambda_1/\cos\theta_B$.  In the right inset a zoom of the $\theta_B=30^0$ is shown with a Lorentzian of width $\lambda_2/(2\cos\theta_B)$.  The figure illustrates how individual collision parameters can be extracted from the interlayer conductivity at finite frequency.}
\end{figure}
As $\theta_B$ is increased, the position of the peak in the conductivity $B=B_\omega/\cos\theta_B$ moves to higher fields.  The peak height also increases.  We noted above that, as long as $\theta_B$ is not too large, the peak in the conductivity where $p\omega_C=\omega$ will be a Lorentzian with width $\lambda_p/(p\cos\theta_B)$.  However once $\theta_B$ becomes big enough that $J_1(k_fc\tan\theta_B)/J_0(k)fc\tan\theta_B)\approx 1$, the peak will no longer be a single Lorentzian and its width will be determined by an average over more than one $\lambda_p$.

In the left panel, a prominent peak occurs for $\theta_B$ as small as $5^0$.  The inset of this panel shows a zoom view of the $\theta_B=5^0$ conductivity curve near its peak compared to a Lorentzian fit with width equal to $\lambda_1/\cos 5^0$.  Clearly, the curve can be well-described by a single Lorentzian and the value of $\lambda_1$ could, in principle, be extracted from this type of analysis.  In the right panel, the peaks are much weaker at small $\theta_B$ because they are associated with the $J_2(\gamma)$ term in Eq. \ref{cond}, which goes to zero rapidly as $\theta_B$ decreases.  This means that curves for larger $\theta_B$ must be considered.  For $\theta_B=30^0$, a peak is evident and can be well fit by a single Lorentzian with width $\lambda_2/(2\cos 30^0)$.  At this angle there will be slight mixing of the $\lambda_3$ and $\lambda_1$ terms since $[J_1(\gamma)/J_0(\gamma)]^2\approx 0.3$ at $\theta_B=30^0$.

\section{Conclusions}

In this article, we calculated AMRO for quasi 2D and 1D metals using an arbitrary elastic collision integral.  The meaning of the effective relaxation rate in AMRO changes depending on the angle between the magnetic field and the current.  When the field is perpendicular to the layers (and parallel to the current), the effective scattering rate is the transport relaxation rate $\lambda_0=\tau_{tr}^{-1}$ while for fields along the layer (perpendicular to the current) the relevant quantity is the total quasiparticle scattering rate $\lambda_\infty=\tau^{-1}$.  For intermediate angles, scattering is dependent on the in-layer momentum dependence of the scattering cross-section, and the associated scattering rate is a weighted average of the intralayer Fourier components of the scattering probability.  We have described methods by which these parameters may be extracted, thus allowing detailed information about the momentum-dependent scattering rate to be obtained.

It is apparent that the simplicity of the models considered here make difficult direct comparisons with data on most systems of interest.  For example, we considered a 2D system that is isotropic in the metallic layers, whereas it is the strong anisotropy in the plane that is the focus of interest in many 2D metals under current investigation.  Nonetheless, the qualitative results presented here should be useful for experimentalists in interpreting their data and, in particular, in comparing AMRO scattering rates with independent determinations.  In future theoretical work the combined effects of strong anisotropy in the layer and strong momentum-dependent scattering probabilities should certainly be investigated.

\section{Acknowledgements}
The authors thank M. P. Kennett for discussions and a critical reading of the manuscript.  This work was supported by an Australian Research Council Discovery Project.

\section{Appendix: Scattering by a random static potential}

To get a clearer picture of the correspondence between the collision parameters and the spatial length scales of the scattering potential, we consider the example of a scattering by a dilute, random distribution of impurities\cite{abrik}.

We write the total potential, due to all impurities, at a given position $\hat{{\bf r}}$ as $U(\hat{{\bf r}})=\sum_{\bf R_i}u(\hat{{\bf r}}-{\bf R}_i)$ where the sum is over the $N_{imp}$ impurities in the sample.  (The position $\hat{{\bf r}}$ is a single-electron operator whereas each impurity position ${\bf R}_i$ is a constant vector since all motion of the heavy impurity is ignored.)  Using Fermi's Golden rule, the probability $W_{{\bf k},{\bf k^{\prime}}}$ that an electron is scattered from ${\bf k}$ to ${\bf k^{\prime}}$ is
\begin{equation}
W_{{\bf k},{\bf k}^{\prime}}=\frac{2\pi}{\hbar}\delta(\epsilon_{\bf k}-\epsilon_{\bf k^{\prime}})|<{\bf k}|U({\hat{\bf r}})|{\bf k}^{\prime}>|^2.
\end{equation}
We average this expression over all impurity configurations by integrating independently each impurity position ${\bf R}_i$ over the sample volume $\Omega$ and keep terms to lowest order in the impurity density $n_{imp}=N_{imp}/\Omega$.  This gives
\begin{equation}
W_{{\bf k},{\bf k}^{\prime}}=\frac{2\pi n_{imp}}{\hbar\Omega}\delta(\epsilon_{\bf k}-\epsilon_{\bf k^{\prime}})u_{\bf q}u_{- {\bf q}}
\end{equation}
where ${\bf q}={\bf k}^{\prime}-{\bf k}$ and $u_{\bf q}=\int d{\bf r} \exp(-i{\bf q}\cdot{\bf r}) u({\bf r})$ is the Fourier transform of the single impurity potential.  Comparing this to Eq. \ref{collI}, we obtain the scattering probability $P({\bf k},{\bf k}^{\prime})$ as
\begin{equation}
P({\bf k},{\bf k^{\prime}})=\frac{n_{imp}}{\hbar(2\pi)^2}u_{\bf q}u_{-{\bf q}}.
\end{equation}
The probability is related to the collision parameters $\lambda_{mn}$ according to Eq. \ref{lambda2}.  If both ${\bf k}$ and ${\bf k}^\prime$ lie on the cylindrical Fermi surface then the probability depends only on $|q_z|$ and $|\phi|$, where $\phi$ is the angle between ${\bf k}$ and ${\bf k}^\prime$, and may be expanded as $P(q_z,\phi)=\sum_{mn}P_{mn}\exp(imq_zc)\exp(in\phi)$.
The collision parameters $\lambda_{mn}$ are
\begin{equation}
\label{collapp}
\lambda_{mn}=\lambda_\infty(1-\frac{P_{mn}}{P_{00}}).
\end{equation}
with
\begin{equation}
\frac{P_{mn}}{P_{00}}=\frac{\int dq_z d\phi \exp(-imq_zc)\exp(-in\phi)|u(q_z,\phi)|^2}{\int dq_z d\phi |u(q_z,\phi)|^2},
\end{equation}
and are thus determined by the Fourier transform of the single-impurity potential.

A simple model of the potential due to a single impurity at the origin in a quasi-2D isotropic system is
\begin{equation}
u(x,y,z)=u_0\exp(-\Delta_z|z|/c)\exp(-2\Delta_\phi k_f\sqrt{x^2+y^2})
\end{equation}
where $u_0$ is a constant.  The range of the potential within a layer is of order $(\Delta_\phi k_f)^{-1}$, where $k_f$ is the Fermi wavevector and $\Delta_\phi$ is a dimensionless scale.  The interlayer range is $c\Delta_z^{-1}$, where $c$ is the interlayer lattice constant and $\Delta_z$ is dimensionless.  We have\cite{zhu04}
\begin{equation}
u(q_z,\phi)=\pi u_0 (ck_f^{-2})\frac{\Delta_z}{\Delta_z^2+(q_zc)^2}\frac{\Delta_\phi}{[\Delta_\phi^2+\sin^2\frac{\phi}{2}]^{3/2}}.
\end{equation}

If the range of the potential is considerably longer than the atomic spacing (so $\Delta_\phi$,$\Delta_z<<1$) then the scattering probability is peaked at zero-momentum transfer and begins to decrease once the intralayer momentum transfer $\phi$ exceeds $\Delta_\phi$ or the interlayer momentum transfer $q_zc$ exceeds $\Delta_z$.
Considering Eq. \ref{collapp}, this implies that $\lambda_{mn}$ will be much smaller than $\lambda_\infty$ if $m<<\Delta^{-1}_z$ and $n<<\Delta^{-1}_{\phi}$ (the arguments of the exponentials are always small and the two terms in the equation nearly cancel in this limit).  However, once $m,n$ approach $\Delta^{-1}_{z}$ and $\Delta^{-1}_\phi$, the second term in Eq. \ref{collapp} will start to drop off and thus the collision parameters $\lambda_{mn}$ will begin to approach $\lambda_{\infty}$.

It thus becomes clear that the scale in $m$, $n$ over which $\lambda_{mn}$ approaches $\lambda_\infty$ gives the spatial range of the scattering potential perpendicular and parallel to the layers respectively.  We found out above that only $m=1$ terms enter the expression for the interlayer conductivity, hence the range of the interlayer potential cannot be obtained in this manner.  The range of the potential within the layers can be obtained, since multiple $\lambda_{1n}$ parameters may be extracted from the interlayer conductivity, as discussed above.

Also, one can at least tell whether the interlayer potential extends over a range of significantly more than one lattice constant by comparing the magnitude of the transport relaxation rate $\lambda_{10}$ to that of the total quasiparticle scattering rate $\lambda_{1\infty}$, both of which can be observed in the interlayer conductivity. For, $\lambda_{10}$ can be significantly smaller than $\lambda_{1\infty}$ only if the factor in the argument of the exponential in Eq. \ref{collapp}, $q_zc$ for $m=1$, is much smaller than unity whenever the scattering probability is non-zero (this is the requirement that the two terms nearly cancel).  This implies that the range of the potential in real space $\Delta_z^{-1}$ is much larger than unity.  Thus the magnitude of the difference between $\lambda_{10}$ and $\lambda_{1\infty}$ gives a clue as to the range of the interlayer scattering potential.

\end{document}